# Entropy controlled fully reversible nanostructure formation of Ge on miscut vicinal Si (001) surfaces

*Christian Grossauer,[1] V. Holy[2] and Gunther Springholz[1]\**

[1] Institut für Halbleiter- und Festkörperphysik, Johannes Kepler University,
A-4040 Linz, Austria.

[2] Department of Condensed Matter Physics, Charles University, 121 16 Praha 2,
and Department of Condensed Matter Physics, Masaryk University, 611 37 Brno,
Czech Republic

ABSTRACT: Entropy effects substantially modify the growth of self-assembled Ge nanostructures on vicinal Si (001) surfaces. As shown by variable temperature scanning tunneling microscopy, this leads to new types of one dimensional nanostructures that are not only tunable in size and shape but can be fully reversible erased and reformed without changes in final sizes and shapes. This unique behavior is caused by the free surface energy renormalization caused by the large step entropy of vicinal surfaces. In thermodynamic equilibrium, this favors the formation of a planar 2D surface at higher temperatures, whereas the nanostructured surface is the preferred low-temperature configuration. Taking the step entropy into account, the critical phase transition temperature is derived by free energy calculations and is shown to scale nearly linearly with the Ge coverage – in excellent agreement with the experiments. Due to self-limitation, the nanowire size is solely controlled by the Ge coverage and vicinal angle, completely independent of the growth or annealing conditions. Thus, highly reproducible nanostructures with tunable geometries are obtained. This opens new avenues for controlled nanostructure formation for practical device applications.

*Corresponding Author, e-mail: gunther.springholz@jku.at

**Introduction**

Self-assembled semiconductor nanostructures produced by strained-layer heteroepitaxy exhibit fascinating features and vast potentials for quantum electronic devices [1–7]. Their formation is based on the fundamental instability of strained layers against surface corrugations or three dimensional (3D) island formation [8–12], which is driven by effective elastic strain relaxation allowed the unconstrained side faces. Due to the highly generic nature of this process, it occurs for a wide range of material systems [13–18] and thus, a rich variety of nanostructures with different sizes and shapes have been obtained, in dependence of coverage, composition and growth conditions [14–24]. Once formed, however, the volumetrically gained elastic energy supersedes the costs in surface and edge formation [10,11,13]. As a result, the total energy monotonically decreases with increasing nanostructure volume [10,12,25], leading to a continuous coarsening and Ostwald ripening [26–28] during growth and annealing. For this reason, strain-induced nanostructure formation has been universally considered to be a *non-reversible* process, meaning that a planar 2D surface can not be regained unless the epitaxial strain is relaxed by misfit dislocations.

In this work, we reveal that due to entropy effects this fundamental reasoning does not apply to nanostructures formed on vicinal surfaces. Entropy effects have been largely neglected in self-assembled growth of nanostructures, i.e., their contribution not been taken into account in the modelling of the total energy of the system [8–10,13]. Here we show that for Ge on vicinal Si surfaces, entropy completely alters the surface evolution. Most unexpectedly, it leads to complete *reversibility* of nanostructure growth, that is, to their complete dissolution above a critical temperature and their reappearance upon cooling. Using high temperature *in vivo* scanning tunneling microscopy (STM) [29,30], we show that this erasure and reformation can be repeated many times, with the final nanostructure sizes and shapes solely controlled by the Ge coverage and vicinal angle, completely *independent* of the initial growth conditions and thermal history. Modelling the total free energy of the system, we identify the large step entropy of vicinal surfaces [31–33] as key factor governing this nanomorphological phase transition. It substantially renormalizes the free surface energy of the vicinal wetting layer, which emerges as the favored high temperature equilibrium phase, whereas the nanostructured surface is the preferred low temperature configuration. The theoretically derived critical transition temperature scales nearly linearly with the Ge coverage - in perfect agreement with our experiments. The generic nature of our model suggests that similar reversible nanostructure growth processes should also exist for other vicinal systems, which provides new pathways for fabrication of perfectly controlled nanostructures for particular device applications.

**Experiments**

Germanium growth on vicinal Si (001) surfaces is studied for a wide range of miscut angles and growth conditions. On vicinal surfaces, the four-fold symmetry of the Si (001) surface is broken and as a result, asymmetric nanostructures are created [14,22,34–36]. In the extreme case, one-dimensional nanowires are formed [34,37] that can be parallel [38,39] or perpendicular [40,41] to the miscut direction. It is noted that similar 1D structures can be obtained by prepattering of the substrate surface [42,43] or by anisotropic incorporation of adatoms, in which case, micrometer long Ge hut nanowires [44,45] can be produced. Due to their favorable in-plane geometry these can be easily integrated into silicon circuits, presenting an attractive platform for realization of *Q*-bits [7,46,47] and Majorana fermions [48].

To monitor the surface evolution during growth and annealing, a multi-chamber molecular beam epitaxy and STM system was employed, allowing surface imaging without breaking ultra-high vacuum conditions [49–51]. In our experiments, Ge was deposited at 1 Å/min on Si substrates with various miscut angles and directions at various temperatures between 450 – 600°C. In this temperature range, Ge/Si intermixing that strongly alters the growth evolution can be neglected [52–54]. Two types of vicinal surfaces were investigated, namely, Si (001) miscut towards (100) by angles α up to 4°, and substrates with miscut towards (110) and α up to 8°. Control experiments on nominally miscut-free singular Si (001) were also performed (see Supplemental Information S1). In the first set of experiments, Ge growth was interrupted at different coverages and the static surface imaged by STM after rapid cooling to room temperature. In the second set, the dynamic evolution of the surface at high temperature was monitored *in vivo* during heating and cooling as a function of temperature.

**Results**

The surface structures of the initial singular or vicinal surfaces are shown by Fig. 1. While the nominally miscut-free Si (001) surface (α < 0.1°) exhibits wide terraces with (2×1) dimer reconstruction occasionally interrupted by isolated monoatomic steps, on vicinal surfaces, a narrow train of steps appear perpendicular to the miscut direction. The average step distance $d_s = h_{ML} \cot \alpha$ decreases with increasing miscut angle, where $h_{ML}$ is the height of the surface steps [1.36Å for Ge (001)]. For the vicinal surfaces with $\alpha = 2°$ and 4° towards (110), the average step distances are thus 3.9 and 1.9 nm, respectively, in good agreement with our STM observations. As shown by Fig. 1c, the miscut steps are parallel or perpendicular to the Si dimers, for which reason they are relatively straight and alternate between the $S_A$ and $S_B$ configuration [29,55,56] and for higher miscut an increasing number of double-layer steps is formed [29,55,57]. For Ge (001), however, step-faceting is less pronounced [57,58]. For vicinal surfaces miscut towards (100), the global step direction is along [010], which is 45° rotated with respect to the Si dimers (see Fig. 1b). For this reason, the steps appear more rough as they are composed of short $S_A$ and $S_B$ segments [58]. It is further noted, that at elevated temperatures, steps increasingly fluctuate and meander due to thermally activated kink formation and diffusion, detachment and reattachments of step atoms [31–33]. This results in increased configurational disorder that is of particular importance for our model described in Sect. 4.

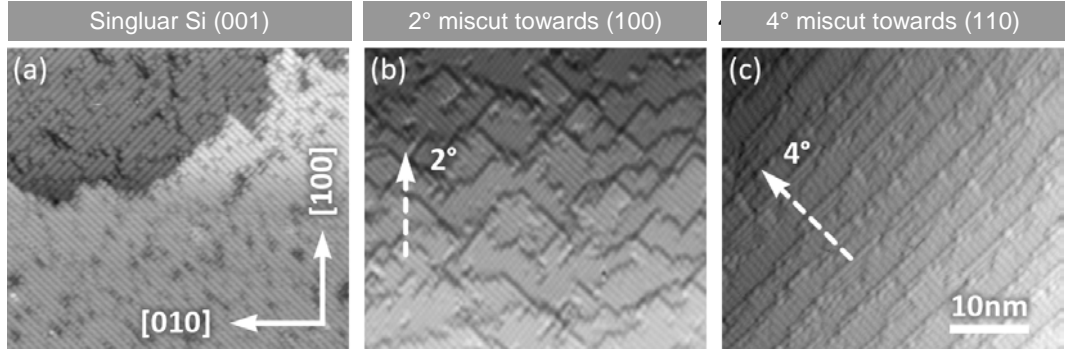

**Figure 1. Step structure** of the initial singular and miscut vicinal Si (001) surfaces as imaged by STM. The miscut angle varies between 0°, 2° and 4° from (a) to (c), respectively, and its direction changes from towards (100) in (b) to towards (110) in (c) as indicated by the dashed arrows. The scale is the same for all images.

For Ge growth, the substrate miscut leads to a substantial modification of the growth process. Whereas on singular Si (001) surfaces at a critical coverage of 6 monolayers (ML) the usual rectangular huts or square based pyramids defined by four low energy {105} side facets are formed (see Supplemental Information S1), on the vicinal substrates a new growth regime emerges, in which 1D nanostructures form much earlier *before* any huts or pyramids appear on the surface. This is shown by Fig. 2 for Ge on 4° miscut Si (001), evidencing that already at 3.5 ML a perfectly facetted 1D nanoripple structure forms (Fig. 2c and d) that seamlessly covers the whole epilayer surface. These new types of structures are confined on one side by a ($\bar{1}$05) facet and on the other side by a (001) facet, while the other three {105} facets of the Ge huts and pyramids are missing (see surface orientation maps (SOM) [14,19,59] depicted as inserts). As shown by Fig. 2(a) and (b), the transition proceeds via initial formation of small step bunches that subsequently merge into small ($\bar{1}$05) microfacets that rapidly expand along [010] perpendicular to the miscut direction, separated by small patches of step-free (001) terraces. After ripples are formed, a second transition takes place in which additional hut-like features nucleate on top of the (001) ripple facets. These huts are confined on all sides by {$\bar{1}$05} facets and resemble those on miscut free Si (001), albeit with highly asymmetric cross section. They are initially sparsely scattered over the ripple facets eventually cover most of the of Ge surface as growth proceeds.

The nanoripples exhibit several intriguing features. First of all, they are formed much earlier (3.5 ML) compared to the huts on singular Si (001), which do not form before a critical thickness of 6 ML (see Supplementary Information S1). Secondly, the ripples seamlessly cover the whole epilayer surface such that nowhere the original 2D wetting layer is exposed.

Last but not least, the whole 2D wetting layers is consumed by the transformation process, i.e., the ripple bases reach down to the Ge/Si interface, where as on miscut free Si (001) the Ge huts always sit on top of the wetting layer. As for the ripple surfaces, the ($\bar{1}$05) ripple facets with characteristic horse-shoe reconstruction [60] are practically step free, as shown by the high resolution STM images presented in the Supplementary Information S2. The (001) ripple facets, on the other hand, appear more rough due to the ($m \times n$) reconstruction of missing vacancy lines and dimer rows, as also seen for the Ge wetting layers on singular Si (001) [29]. Nevertheless, the (001) facts are still practically free of miscut steps because the whole miscut is accommodated by the ($\bar{1}$05) ripple facets.

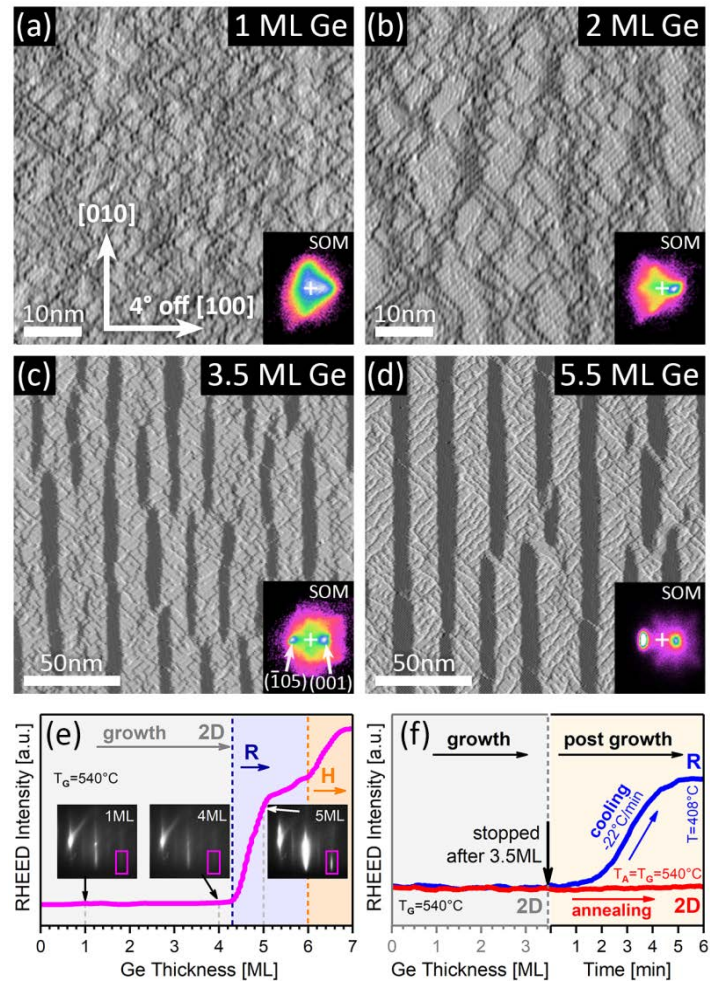

**Figure 2. Nanoripple formation during Ge growth on 4° vicinal Si (001),** as seen by STM (a-d) at Ge coverages increasing from 1 to 5.5 ML. The nanoripples are perpendicular to the miscut direction and consist of alternating ($\bar{1}$05) and (001) facets (*see* surface orientation maps (SOM) shown as inserts). The STM images were recorded at room temperature after cooling from $T_{Growth}$ = 540°C. Note the different scales. (e) RHEED intensity evolution during growth, indicating a critical coverage of 4.3 ML for ripple formation. At 6 ML, an additional kink appears due to hut nucleation. Representative RHEED patterns are shown as inserts. Panel (f) displays the intensity evolution *after* Ge growth of 3.5 ML. When the temperature is kept constant at 540°C (red trace), no change in intensity, i.e., no ripple formation occurs. In contrast, when the temperature is ramped down to room temperature (blue trace), a strong intensity increase takes place. This signifies, that the nanoripples at this coverage are actually formed not during but rather after growth during the cool down process.

The exact onset of 1D nanoripple formation was determined by measuring the intensity evolution of RHEED diffraction spots *in situ* during growth. The result is displayed in Fig. 2e, evidencing that for the given growth temperature of 540°C the onset occurs at 4.3 ML – in surprising contradictions to what we see by STM after cooling to room temperature, where ripples are seen already at a considerably lower coverages of 3 ML. This signifies that a substantial surface evolution actually takes place during the cool down process. To clarify this puzzling contradiction, we have tracked the RHEED evolution after growth termination. The result is displayed in Fig. 2f, where in one case after growth interruption at 3.5 ML the temperature was either kept constant at 540°C (red line) or in the other case was ramped down at by -22°C/min to room temperature (blue line). In the first experiment, the diffracted intensity remains completely constant even for prolonged time periods of hours without any sign of nanostructure formation. On the contrary, when the sample temperature is ramped down to room temperature, the diffracted intensity rapidly increases, signifying ripple formation once the temperature falls below 500°C. This means that at lower coverages the nanoripples are actually not formed *during* but *after* growth during the cool down process.

**Reversible nanoripple formation**

To shed light on this unexpected phenomenon, we follow the surface evolution as a function of temperature *in vivo* using high temperature STM as shown by Fig. 3. To this end, we start with well-developed nanoripples formed at 4.6 ML Ge coverage after the 1$^{st}$ cool down process (Fig. 3a). The temperature was then slowly ramped in steps up to 580°C and then back to room temperature, while STM images were continuously recorded. The resulting snap shots extracted from this STM "movie" are displayed in Fig. 3b-f, with the color coding indicating the temperature at which each image was recorded (see scale on the right hand side). Starting from well-developed nanoripples, heating up to 500°C evidently does not have any effect on the stable ripple structure. Above 500°C, however, the ripples gradually dissolve, such that at 580°C a completely flat 2D surface is regained (Fig. 3d). Conversely, upon lowering again the temperature, the ripples reappear and below 500°C a fully facetted ripple structure is regained. Most remarkably, we find that this dissolution and reformation process can be repeated many times without any appreciable change in the final ripple structure. This is evidenced by the STM image of Fig. 3f, which shows the ripple surface after *five* annealing/cooling cycles. Evidently, is undiscernible from that directly obtained after growth (Fig. 3a), which means that ripple formation is not only *fully reversible*, but even more, that the final structure is *completely independent* of the thermal history and growth conditions. Thus, the ripples represent a stable equilibrium morphology formed below certain critical temperature, whereas above the 2D surface is the equilibrium state of the system. This is the key result of our experiments.

We also find that the critical temperature of nanomorphological transition strongly depends on the Ge coverage. To this end, we have performed a series of annealing experiments using *in situ* RHEED as shown in Fig. 3g. Clearly, for all Ge coverages between 3 to 5 ML, a fully reversible 3D/2D transition occurs, characterized by a high intensity for the low temperature ripple phase and a low intensity for the 2D phase. For each coverage, completely *identical* traces are observed for repeated annealing cycles that plotted on top of each other in Fig. 3g. This underlines the perfect reproducibility of the transition and confirms its thermodynamic origin. Most importantly, the transition continuously shifts to higher temperatures when the Ge coverage increases, *i.e.*, it occurs at around 465°C for 3 ML, whereas at 5 ML is shifted

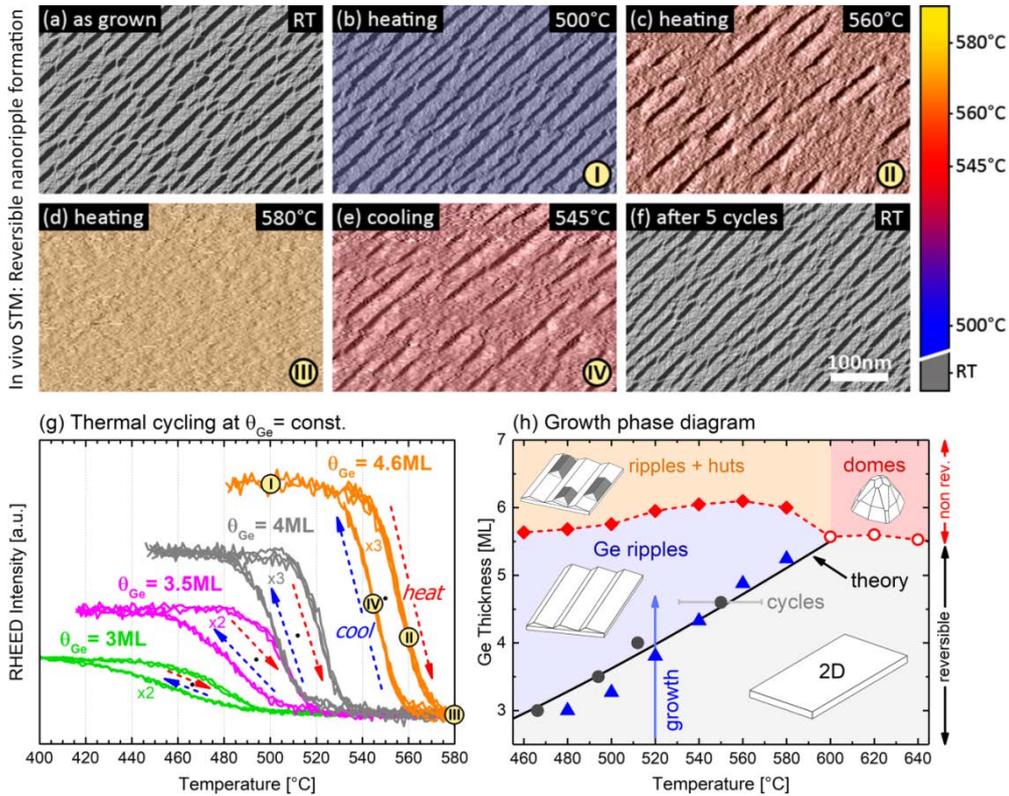

**Figure 3. Fully reversible Ge nanoripple formation on 4° vicinal Si (001)** miscut towards (100) revealed by *in vivo* high-temperature STM and RHEED. The sequence of STM images (a-f) was recorded during heating of 4.6 ML Ge from room temperature to 500°C, 560°C and 580°C, and then cooling back to 545°C and finally room temperature, respectively. Complete ripple dissolution takes place at 580°C and perfect ripples are re-formed when the temperatures is lowered back below 520°C. No difference between the initial and final ripple structure is seen after the complete annealing cycle (f). RHEED intensity curves recorded during multiple heating/cooling cycles as a function of temperature ($r_T = \pm 12$°C/min) are shown in (g) for different Ge coverages between 3 to 4.6 ML, demonstrating how the transition temperature increases with Ge thickness. For each coverage, several consecutive heating/cooling cycles (×2 and ×3) are plotted on top of each other. The resulting equilibrium surface phase diagram is shown in (h). The black dots represent the 2D/3D transition temperature derived from annealing experiments, the blue triangles the ripple onset during growth (Fig. 1g). The solid black line represents the critical transition temperature derived by total free energy calculation as described by Eq. 8. The red diamonds indicate the transition from ripples to huts, and the red circles the formation of multi-facetted domes at higher temperature and higher coverages.

Ge coverage increases, *i.e.*, it occurs at around 465°C for 3 ML, whereas at 5 ML is shifted to 570°C. In addition, a hysteresis between heating and cooling is observed, as is typical for first order phase transitions. Its width depends on the heating/cooling rates and narrows down to a few degrees when they are reduced to ±1°C/min. Therefore, it is assigned to kinetic effects, and accordingly it is wider at lower coverages, i.e., lower transition temperatures. As we do not observe any change for subsequent annealing cycles, Si/Ge intermixing or interdiffusion is negligible for the used annealing conditions.

The complete data set allows us to assemble a complete equilibrium phase diagram of Ge on the vicinal Si surface as presented in Fig. 3h, indicating the equilibrium surface morphologies as a function of temperature and Ge coverage. Clearly, for coverages below 5.5 ML, the surface is either in the ripple (blue region) or 2D phase (grey region). The critical temperatures $T_C$ for this transition derived from the annealing experiments are represented by the black dots, and the critical coverage for the ripple onset during growth by the blue triangles. The critical temperature clearly rises nearly linearly when the Ge coverage increases. Also indicated is the onset of hut nucleation on top of the ripples (red diamonds), occurring at about 5.5 ML independently of temperature. The regime where pyramids and domes are formed at temperatures higher than 600°C is also indicated. At these conditions, however, strong Si/Ge intermixing sets in [17,52–54]. Thus, the full reversibility of the nano-morphological transition is lost, *i.e.*, dome islands cannot be dissolved by annealing, but only coarsen, just as the domes on miscut-free singular Si (001) surfaces [61–64]. Although, cooling down such high-temperature annealed surface again leads to ripple formation in the area between the domes, these ripples do no longer attain the same size and uniformity of those for low temperature annealed samples the wetting layer thickness is diminished by the material transferred into the domes (see Supplementary Information S3).

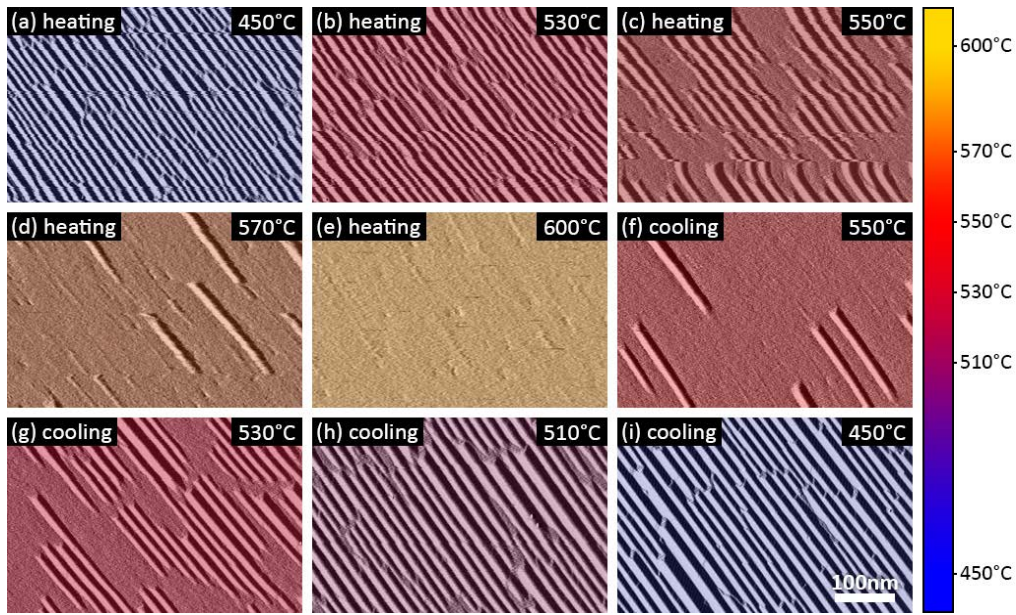

**Figure 4. Fully reversible Ge nanoripple formation on 8° vicinal Si (100) with miscut towards (110)**, as seen by *in vivo* high-temperature STM images recorded during temperature ramping from 450° to 600°C (a-e) and cooling back to 450°C (g-i). The Ge coverage was 4.7 ML.

Reversible nanoripple formation is not unique for a particular miscut angle but is observed for a wide range of vicinal angles and directions. This is proven by analogous experiments performed for 2° - 8° miscut vicinal Si (001) surfaces, where in all cases, 1D Ge nanoripples are formed, albeit with different ripple geometries and facet angles. In each case, we find that annealing above a certain temperature leads to ripple dissolution and their reformation when cooled back below $T_C$. This is demonstrated by the sequence of *in vivo* STM images recorded for Ge ripples formed on 8° vicinal Si (001) during annealing between 450° to 600°C. Evidently, dissolution/reformation occurs in this case at similar temperatures as observed for 4° vicinal Si shown in Fig. 3. This is quite remarkable, because the ripples on 8° miscut Si exhibit a quite different ripple geometry (see Supplementary Information S2) with symmetric {105} facets on both sides as shown in Refs. [36,38]. Moreover, we find in this case a similar increase of $T_C$ from 460° to 560°C when the Ge coverages increases. This highlights the completely generic nature of the reversible process on vicinal surfaces that is solely controlled by the Ge coverage and temperature.

**Scaling of the nanowire period**

A particular feature of nanowire formation is the quasi-deterministic control of size and period obtained by varying the Ge coverage. This arises from the fact that ripple formation is a self-limiting process that stops once the Si/Ge interface is reached. Accordingly, the total ripple volume equals the amount of Ge deposited, which means that the ripple height $h_R$ is simply twice the deposited Ge thickness. For a fixed ripple geometry, the equilibrium ripple period $p_R$ thus scales linearly with the Ge coverage

$$p_R = \lambda_\alpha \theta_{Ge} \qquad (1)$$

where the scaling factor $\lambda_\alpha$ is solely determined by aspect ratio of the ripples. To test this prediction, we plot in Fig. 5a the equilibrium ripple periods determined for a wide range of coverages and miscut angles (see Supplementary Information S4), evidencing a linear scaling of the period for each investigated miscut angle, indicating that the above model assumptions hold. The slope $\partial p_R / \partial \theta_{Ge}$, *i.e.*, the value of the scaling parameter $\lambda_\alpha$ differs for the different miscut angles due to the different resulting ripple geometries, and can be easily derived as

$$\lambda_\alpha = (\cot \alpha + \cot \beta)/2h_{ML} \qquad (2)$$

where $\alpha$ and $\beta$ are the inclinations of the ripple facets to the vicinal substrate surface.

For vicinal surfaces miscut towards (110) (*cf.* Fig. 2), the ripples consists of alternating (001) and (105) facets and thus, $\alpha$ is simply the substrate miscut and $\beta = 11.3° - \alpha$, where 11.3° is the angle between (105) and (001). The resulting dependence of $\lambda_\alpha$ versus miscut angle is shown in Fig. 5b, indicating that $\lambda_\alpha$ rapidly decreases but reaches a minimal value of $\lambda_\alpha$ = 5.6 nm/ML at $\alpha$ = 5.65°, where the (105)/(001) ripples are symmetric ($\alpha = \beta$) and exhibit the highest aspect ratio. For larger miscut angles, the asymmetry and accordingly, $\lambda_\alpha$ increases. The experimental values (full circles) obtained from the fits of the experimental data nicely fall on the theoretical line. For the vicinal surface 8° with miscut towards (100), the ripple facets are inclined by $\alpha = \beta$ = 7.97°, giving a scaling factor of $\lambda_\alpha$ = 4 nm/ML (dashed line in Fig. 5a), which is again in good agreement with the experimental. This underlines that the model is valid for a wide range of surface orientations.

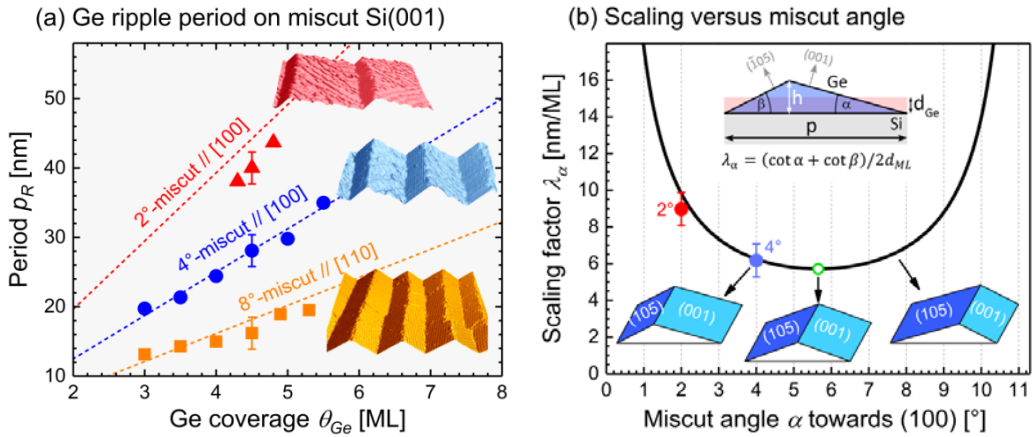

**Figure 5. Scaling of the equilibrium ripple period** $p_R$ as a function of Ge coverage for three different vicinal Si surfaces. Red triangles and blue dots corresponds to Ge on substrates with 2° and 4° miscut towards (100), orange squares to those on 8° Si miscut towards (110). The data was derived from autocorrelation analysis of STM images recorded after slow cooling through the 2D/3D transition (see Supplemental Information S4). The dashed lines represent the ripple periods predicted by Eqs. 1 and 2, and high resolution STM are shown as inserts to illustrate the different ripple geometries. Panel (b) shows how the scaling parameter $\lambda_\alpha = p_R/\theta_{Ge}$ depends on the miscut angle $\alpha$. The symbols represent the experimental values.

## The critical transition temperature

To clarify the origin on the reversible nanoripple transition, we model the temperature dependence of the free energy difference between the ripple and 2D surface. For strained heteroepitaxial systems, this difference is given by [37,38]

$$\Delta E_{tot} = E_R - E_{2D} = \Delta E_{strain} + \Delta E_{surf} + \Delta E_{edge} \quad (3)$$

where $\Delta E_{strain}$ (< 0) is the elastic energy relaxed by the ripples relative to that of the 2D surface, $\Delta E_{surf}$ the difference in free surface energy and $\Delta E_{edge}$ the energy associated to the edges formed between the ripples facets. When $\Delta E_{tot}$ < 0, the ripple phase is favored, whereas for $\Delta E_{tot}$ > 0, the 2D surface is the stable configuration. $\Delta E_{strain}$ scales linearly with the Ge coverage because the relaxed elastic energy per unit volume $\rho_{el}$ is fixed for a given ripple geometry [37,38], and because the ripple volume is equal to the volume of the wetting layer. $\Delta E_{surf}$ on the contrary is independent of Ge coverage because the ripple geometry is fixed and the ripples seamlessly cover the whole epilayer surface. For the sake of simplicity, here we neglect any changes of surface energy as a function of distance from the Si/Ge interface, that can be only obtained by density functional theory calculations [65–67]. Last but not least, $\Delta E_{edge}$ scales inversely to the Ge coverage because there are two edges per ripple and the ripple width increases linearly with the coverage (Fig. 5). Per unit area, the total energy difference thus reads as

$$\Delta E_{tot}/A = -\rho_{el}h_{ML}\,\theta_{Ge} + \gamma_R^* - \gamma_{2D} + \frac{e_{edge}}{\lambda_\alpha\,\theta_{Ge}} \qquad (4)$$

where $\gamma_R^* = 2h_0(\gamma_{001}\csc\alpha + \gamma_{105}\csc\beta)/\lambda_\alpha$ is the projected surface energy of the (001) and (105) ripple facets, $\gamma_{2D}$ the surface energy of the vicinal wetting layer and $e_{edge}$ the edge energy per unit ripple length.

At sufficiently large coverages, the energy difference is dominated by the 1$^{st}$ term of Eq. 4, i.e., the elastic energy relaxation. This means that without further corrections there is *one* unique critical coverage $\theta_c$ beyond which the 3D ripple phase is favored over the 2D phase. To explain the reversibility of the ripple transition, therefore the temperature dependence of the factors in Eq. 4 need to be taken into account. For a fixed ripple shape and composition, the relaxed elastic energy density $\rho_{el}$ is temperature independent - not considering the minute changes in lattice parameters and elastic constants. The edge energies are also unlikely to change much, provided that the atomic configuration at the facet edges remains unchanged. This leaves surface energies as main source for the free energy renormalization. Specifically, as the reversibility is clearly linked to the vicinality of the substrate surface, we identify the step entropy of the highly stepped wetting layer surface as the key factor for this process.

At elevated temperatures surface steps start to meander and fluctuate via thermal kink formation and annihilation [31,32,68–70]. The corresponding configurational disorder reduces the free surface energy of the vicinal wetting layer according to

$$\gamma_{2D,\alpha}(T) = \gamma_{2D,0} - n_{s,\alpha}T\,S_s \qquad (5)$$

where $n_{s,\alpha} = \tan\alpha/h_{ML}$ is the step density of the vicinal surface, $S_s$ the step entropy and $\gamma_{2D,0}$ the surface energy at zero Kelvin. The magnitude of the entropy contribution is directly proportional to step density. Thus, it is particularly large for the high-miscut wetting layer surface, but negligible for the singular, practically step free ripple facets, where $\gamma_R$ can be assumed to be constant. This means that the surface energy of the vicinal wetting layer substantially decreases as the temperature increases. At high temperatures, this shifts the free energy balance shifts towards the 2D phase, eventually inducing a ripple-to-2D phase transition once a critical temperature is reached.

For Ge (001), step meandering is significant above the freeze-in temperature $T_F = 575K$ [70], as applies for our current experiments. As detailed by Zandvliet [70], the step entropy of Ge surface steps along [010] consists of the sum of step meandering $S_m$ and vibrational entropy $S_{vib}$ and can be written as

$$S_s = S_m + S_{vib} = \frac{k_B}{a_0}\ln(1 + e^{-\delta/2k_BT}) - \frac{3k_B}{2\,a_0}\ln\left(\frac{1-e^{-\Theta_S/T}}{1-e^{-\Theta_T/T}}\right) \quad (6)$$

In this equation, $\delta = -5$meV is the next-nearest step atom interaction energy, which for Ge (001) is very small [70]. This means that $S_m$ is practically constant above 300K. Likewise, the vibrational entropy (second term) arising from the reduced coordination of the step atoms, determined by the Debye temperatures $\Theta_S = 264K$, respectively, $\Theta_T$ 341 K [70], is also only very weakly temperature dependent. Thus, for the temperature range pertaining to our experiments (600 – 873K), the step entropy can be approximated by $S_s^* \cong k_B/a_0$ (see Supplementary Information S5). Here we do not include step-step interactions as well as possible faceting the steps [70], which would somewhat modify the terms in Eq. 5, but would not affect the generic behavior of our model. Combining all terms, the total free energy difference per unit area, including step entropy, now reads as

$$\Delta E_{tot}(T, \theta_{Ge})/A \cong -\rho_{el}h_{ML}\theta_{Ge} + \Delta\gamma_{s,0} + n_{s,\alpha}S_sT + \frac{e_{edge}}{\lambda_0}\frac{1}{\theta_{Ge}} \quad (7)$$

where $\Delta\gamma_{s,0} = \gamma_{R,0}^* - \gamma_{2D,0}$ is the surface energy difference without entropy corrections.

At the 2D – 3D phase transition, the energy of the ripples and the 2D wetting layer are equal, i.e., $\Delta E_{tot} = 0$. This yields the transition temperature $T_c$ as

$$T_c(\alpha, \theta_{Ge}) = \frac{1}{n_{s,\alpha}S_s}\left(-\Delta\gamma_{s,0} + \rho_{el}h_{ML}\theta_{Ge} - \frac{e_{edge}}{\lambda_\alpha}\theta_{Ge}^{-1}\right) \quad (8)$$

The critical temperature $T_c \sim 1/n_{s,\alpha}S_s$ scales inversely to the step density on the vicinal wetting layer, which means that $T_c$ is sufficiently low only for high miscut surfaces. This explains why a reversible 2D - 3D transition has not been observed on singular Si (001). For the 4° miscut Si (001) surface (Fig. 3), the step density is $n_s = 0.05$ Å$^{-1}$ and the elastic energy relaxation is $\rho_{el} = 0.024$meV/Å$^3$ as obtained by solving the surface stress equations in the shallow slope approximation [71,72] (see Supplementary Information S5). Using $S_s^* \cong k_B/a_0$ and $e_{edge} = 8$ meV/Å as suggested by Retford et al. [73], we can nicely reproduce the $T_c(\theta_{Ge})$ dependence seen in our experiments by our calculations, represented by the black line in Fig. 3g. Thus, the reversible transition is well explained by step entropy effects. From the fit we obtain $\Delta\gamma_{s,0} = -0.5$ meV/Å$^2$ between the ripple and 2D phase for α = 4°. Most importantly, the existence of the reversible 2D/3D transition does not rely on particular details of the step structure of the vicinal surface, as the step entropy will always reduce its free surface energy when the temperature increases. This is the basis for this effect and is the reason why we observe this transition for a wide range of vicinal surfaces.

Our model also explains why the reversibility is lost once higher aspect ratio structures such as pyramids or domes are formed on the surface, because in this case, the elastic energy relaxation $\rho_{el}$ strongly increases. According to Eq. 8, this shifts $T_c$ upward into the regime where coarsening [62,74] as well as irreversible Si/Ge intermixing sets in [17,52–54], which completely alters the evolution of the system. It is also noted that with changing miscut angle, not only the step density $n_s$, but also the surface energy difference $\Delta\gamma_{s,0}$ and elastic energy relaxation $\rho_{el}$ will change due to the changing ripple shape, which will affect the transition temperatures. These changes however tend to cancel each other in Eq. 8, for which reason the $T_c$´s remain in a similar temperature range for a wide range of miscut angles as we experimentally observe. Most importantly, Eq. 8 reveals that for sufficiently large $\theta_{Ge}$, the critical $T_c$ scales nearly linearly with Ge coverage in excellent agreement with our experiments (see Fig. 5g). Together with the deterministic control of the ripple period, this distinguishes the reversible ripple formation from defaceting transitions of bulk vicinal surfaces [75–77], where the transition not only occurs at significantly higher temperatures, but also the critical temperatures do not depend on strain or coverage, and the facet sizes continuously grow in time because of the absence of a self-limiting growth factor [75–77].

**Conclusion**

In conclusion, we have demonstrated that entropy effects can substantially modify the growth of self-assembled nanostructures on non-singular vicinal substrate surfaces. For the prototypical Ge/Si system, this leads to a full reversibility of nanostructure formation, in which these nanostructures can be repeatedly erased and reformed many times with a final structure that is completely independent of the initial growth conditions and annealing history. The reversibility is caused by the free energy renormalization caused by the large step entropy of vicinal surfaces, which favors the 2D surface at higher temperatures, whereas the nanostructured surface is the preferred low-temperature configuration. Because nanoripple growth is self-limited by the Si/Ge interface, their size can be deterministically controlled by the Ge coverage and vicinal angle. As a result, highly reproducible nanostructures with tunable geometries can be obtained. Last but not least, our results reveal an astonishingly large surface mass transport even after growth termination, leading in the extreme case to a complete restructuring of surface morphology during the cool down process. This means that post mortem images of epitaxial surfaces does not always represent the actual growth morphologies as is usually presumed, which is an important fact that needs to be taken into account in the analysis of self-assembled growth processes. The generic nature of our model and its validity for a wide range of miscut angles suggests that similar effects should also take place for other vicinal systems. This opens up new opportunities for realization of novel nanostructures for device applications.

**Acknowledgments**

We thank Istvan Daruka for developing the original free energy model taking the step entropy into account. This work was supported by the Austrian Science Funds, project P28185-N27.

**Supplementary Information**

The supplemental information contains (i) the comparison to Ge growth on singular Si (001) substrates, (ii) high resolution STM images to reveal the atomic structure of Ge ripples on 4° and 8° vicinal Si (001), (iii) the description dome formation during high temperature annealing, (iv) the determination of the ripple size and periodicity, and (v) details on the step entropy and energetics of phase transition described by our model.